\documentclass[12pt,a4paper]{article}
\usepackage{authblk}
\usepackage{tikz-cd}
\usepackage[english]{babel}
\usepackage[latin1]{inputenc}
\usepackage{lmodern}
\usepackage{verbatim}
\usepackage{amsfonts}
\usepackage{amsmath}
\usepackage[T1]{fontenc}
\usepackage{color}
\usepackage[normalem]{ulem}
\usepackage{natbib}
\usepackage{url}
\usepackage[bookmarksnumbered=true, colorlinks=true, citecolor=blue]{hyperref}
\date{}

\title{How Humean is Bohumianism?}
\author{Tomasz Bigaj}
\affil{University of Warsaw, Institute of Philosophy\\ Krakowskie Przedmie\'scie 3\\ 00-927 Warsaw}
\author{Antonio Vassallo}
\affil{International Center for Formal Ontology\\Faculty of Administration and Social Sciences\\Warsaw University of Technology\\Plac Politechniki 1\\ 00-661 Warsaw}

\begin{document}

\maketitle
\begin{center}
Accepted for publication in \emph{Foundations of Physics}
\end{center}
\pdfbookmark[1]{Abstract}{abstract}
\begin{abstract}
An important part of the influential Humean doctrine in philosophy is the \emph{supervenience} principle (sometimes referred to as the principle of separability). This principle asserts that the complete state of the world supervenes on the intrinsic properties of its most fundamental components and their spatiotemporal relations (the so-called Humean mosaic). There are well-known arguments in the literature purporting to show that in quantum mechanics the Humean supervenience principle is violated, due to the existence of entangled states. Recently, however, arguments have been presented to the effect that the supervenience principle can be defended in Bohmian mechanics. The key element of this strategy lies in the observation that according to Bohmian mechanics the fundamental facts about particles are facts about their spatial locations, and moreover, for any proper subsystem of the world its state may non-trivially depend on the spatial configuration of the rest of the universe. Thus quantum-mechanical states of subsystems do not represent their intrinsic properties but rather characterize their relations with the environment. In this paper we point out the worry that this Bohmian strategy --known as \emph{Bohumianism}-- saves the letter but not the spirit of the Humean doctrine of supervenience, since it prima facie violates another seemingly important Humean principle, which we call \emph{Strong Supervenience} and whose denial implies the existence of necessary connections among distinct individuals. We argue that the best defense for Bohumians is to question the fundamental existence of complex physical systems and their states by treating any reference to them as a convenient description of the underlying collection of Bohmian particles. We consider several pros and cons of this strategy.
\end{abstract}

\section{Introduction}

The main purpose of this paper is to critically evaluate the latest attempts to reconcile modern developments in quantum physics with the broad philosophical doctrine of Humeanism. It is no secret that the worldview emerging from quantum mechanics turns out to be rather hostile to many venerated philosophical views, such as for instance determinism. The doctrine of Humeanism, with its insistence on separability, non-modality and reductionism, seems to be another prime target for an assault inspired by quantum discoveries. And yet, given the persistent popularity of Humeanism among contemporary philosophers of science, no wonder that this assault is met with powerful resistance. One recent episode in an ongoing battle between various philosophical approaches to quantum mechanics started out with a vigorous attack on Humeanism at the hands of Tim Maudlin in his provocatively titled piece ``Why be Humean?'' \citep{478}. It didn't take long before defenders of the ``greatest denier of necessary connections'' responded to this challenge in various ways. A particularly interesting response has been based on a popular in philosophical circles interpretation of quantum mechanics known as Bohmian mechanics \citep{382,477,580,482}. In our contribution to this debate we will try to identify some weak points of this strategy, and we will suggest, on behalf of the Bohmians, how to strengthen their defensive position.

The plan of this article is as follows. Section \ref{intr} is mainly expository, devoted to the presentation of the main assumptions and arguments in Maudlin's original polemic with Humeanism, as well as the general strategy of defense against Maudlin's attack. Here we will introduce two pillars of the Humean doctrine: the rejection of modal facts and the elimination of necessary connections between separate entities, and we will present the well-known Humean principles of Physical Statism and Separability. Reconstructing Maudlin's criticism of Separability based on the example of entangled states (singlet- and triplet-spin states), we will formulate an equivalent version of Separability that we call Supervenience, and we will show, following Elizabeth Miller, how Supervenience can be defended by denying that the singlet and triplet-states are intrinsic to the quantum systems. In section \ref{bsabm} we will explore the physical details of Bohmian mechanics relevant to the above-mentioned strategy of defending Humeanism. We will stress the nomological character of the universal wave function (interpreted according to the Best System approach), and we will introduce the concept of an effective wave function that enables us to make a distinction between physical systems and their environments while still admitting that the states of these systems are not intrinsic to them.

Section \ref{se3} is central to the article. In it we present our main critique of the Bohmian defense of Humeanism, which is that Separability (or Supervenience) by itself does not guarantee that no necessary connections between separate entities will be present. Supervenience ensures the lack of necessary connections between fundamental objects, but does not exclude the possibility that these connections may emerge at higher levels of complexity. To clarify this further, we spell out the conditions of Strong Supervenience and Binary Separability that in our opinion should constitute parts of the broader Humean doctrine, and we argue that Bohmian mechanics refutes them. In section \ref{se4} we discuss one possible way for Bohmians to continue their commitment to Humeanism by denying the objective existence of complex entities on a par with fundamental objects. We explain why this move does not imply that objects of our experiences are mere illusions or arbitrary constructs. We end the article with a suggestion for Bohmians to bite the bullet and go ``Democritean'', i.e. accept that only fundamental particles (``atoms'') exist in the most literal sense of the word.

\section{Humeanism and Quantum Mechanics}\label{intr}
 
The modern version of Humeanism is based on two intuitions, both of which can be traced back to David Hume himself. One of them is skepticism regarding irreducible modal facts: facts involving necessity, possibility, dispositions, powers, counterfactuals etc. The other intuition suggests that there can be no necessary connections between separate entities.\footnote{We are not claiming that these two intuitions are independent from one another. In particular, it may be argued that rejecting irreducible modal facts, as per the first intuition, eliminates necessary connections between distinct entities. On the other hand, it is at least feasible to reject irreducible modal facts as parts of the world, and still insist that there are some necessary connections which present themselves in the fact that certain combinations of fundamental, non-modal properties are not admissible.} The first intuition can be expressed in the statement that all facts about the world are exhausted in its total state that tells us what the world \emph{is}, as opposed to what it \emph{could}, or \emph{would}, or \emph{ought} to be. All purported modal facts, encompassed for instance in laws or tendencies or dispositions or whatever, are to be ultimately reducible to the total state of the world (in the sense that there could be no difference in modal facts without the underlying difference in the total state).

The non-existence of necessary connections between separate individuals gives us more insight into the internal composition of the total state of the world according to Humeanism. Roughly speaking, the state of the world should be decomposable into the states of separate individuals plus their spatiotemporal arrangements. The so-called Humean mosaic, consisting of the assignment of intrinsic, natural and non-modal properties to individual objects, together with the spatiotemporal arrangement of these objects, should be sufficient to fix the state of the entire world. Admitting any facts beyond what is included in the Humean mosaic would amount to the acceptance of fundamental correlations between distinct entities that are not reducible to their intrinsic properties and spatiotemporal relations, and hence would infringe the second Humean intuition. 

Tim Maudlin's verbalization of Lewis's variant of Humeanism has become the gold standard in the literature. He spells out two principles he calls \emph{Physical Statism} and \emph{Separability} as follows \citep[][p. 51]{478}:

\begin{quote}
(Physical Statism) All facts about a world, including modal and nomological facts, are determined by its total physical state.

(Separability) The complete physical state of the world is determined by (supervenes on) the intrinsic physical state of each spacetime point (or each point-like object) and the spatiotemporal relations between those points.
\end{quote}

As \citet[][p. 74]{482} point out, Physical Statism and Separability jointly imply that the world is fundamentally non-modal, and that there are no necessary connections at the fundamental level. The last supposition is secured by Separability, since Separability ensures that fundamental objects are characterized by intrinsic states, and intrinsic properties by definition are possessed by an entity independently of anything else in the world. It is worth stressing that the supervenience basis described in Maudlin's Separability consists of fundamental entities (points or point-like entities) that cannot be further decomposed into smaller constituents. This is a rather important restriction, since from Separability understood in such a way it does not immediately follow that an analogous principle will hold for non-fundamental objects and their properties. In other words, it may still be the case that the total state of the world is not reducible to the Humean mosaic consisting of larger, non-fundamental individuals (for instance middle-size objects of our everyday experience).

Maudlin argues that Separability comes under severe fire from the development of quantum physics. Here entangled states provide the required ammunition. An entangled state of a two-particle system is a state that cannot be written as a product of two one-particle states. It is possible to find two distinct entangled states such that both give the exact same predictions regarding the behavior of individual particles taken separately (they produce the same reduced states for each particle), and yet they differ with respect to the global measurable features of the entire system. The example given by Maudlin uses the singlet and triplet spin-states of two electrons. Both states give identical statistical predictions regarding individual electrons: the probability of obtaining any result (up or down) for spin in any chosen direction equals $50\%$. And yet when the system is in the singlet spin state the results of measurements of spin in the same direction on both particles are always anticorrelated (they have opposite values), while in the case of the triplet state we can choose a direction of spin measurement in which both outcomes must be the same. Thus the quantum state of the whole two-particle system does not supervene on the intrinsic properties of the separate particles and their spatiotemporal locations. 

However, the argument from quantum entanglement does not directly affect Maudlin's Separability, as the latter principle concerns the world as a whole and not small isolated systems such as a two-electron system. In order to question Separability we need an additional assumption that would enable us to get from the violation of Humean supervenience at a local level to the global violation. The missing link was identified by Elizabeth Miller \citep{382} as the claim that the quantum state of a two-electron system forms its \emph{intrinsic} property.\footnote{We do not wish to enter the intricate debate on the exact meaning of ``intrinsic properties''. For our purposes it is sufficient to accept the following partial characteristic: if an intrinsic property of a system $X$ supervenes on any property of an object $Y$, $Y$ has to be a spatiotemporal part of $X$.} From this assumption we can now derive that the singlet and triplet states of two electrons cannot supervene on properties of objects other than the electrons themselves, since this would contravene the intrinsic character of the two-particle states. Thus neither singlet nor triplet states supervene on \emph{any} properties of the fundamental elements of the world (whether or not these elements are parts of the considered system) and their spatiotemporal arrangements. And because the total state of the world must contain the states of its components, Separability is violated.

Proceeding slightly more formally, we can first introduce a new principle that can be simply called \emph{Supervenience}:

\begin{quote}
(Supervenience) For any physical system $S$, the intrinsic properties of $S$ supervene on the intrinsic physical states and spatiotemporal relations between points and pointlike objects that are parts of $S$.\footnote{Miller calls the negation of Supervenience \emph{Metaphysical holism} \citep[][p. 568]{382}.}
\end{quote}

Given some reasonable assumptions, it can be proven that Supervenience is in fact equivalent to Separability. Because trivially all the properties of the entire world are intrinsic to it, Separability follows from Supervenience (if we assume that the world constitutes a physical system, which seems plausible). The entailment in the opposite direction can be shown as follows. If Supervenience were false, some intrinsic properties of a particular system $S$ would not supervene on the Humean mosaic of fundamental, point-like objects \emph{that are parts} of $S$. But intrinsic properties of system $S$ cannot depend on any object spatiotemporally external from $S$, hence the intrinsic properties of $S$ do not supervene on the fundamental properties of point-like objects at all. Provided that the intrinsic properties of any physical system are parts of the total state of the world, this means that Separability is false. 

It is clear that the example of the singlet and triplet states directly violates Supervenience, if only these states qualify as \emph{intrinsic} properties of appropriate systems. But here one escape route immediately opens up. We can namely attempt to save Supervenience (and Separability) by denying that the singlet and triplet states represent genuine intrinsic properties of physical systems. This is precisely the strategy adopted by Miller in her attempt to defend Humeanism against Maudlin's argument. If we can argue that the singlet or triplet states of a system of two particles depend non-trivially on the properties of objects extrinsic from this system, Maudlin's argument can be repelled.\footnote{Another strategy of how to deal with Maudlin's challenge to Humeanism has been considered by George Darby \citep{579}. The idea is roughly to include in the supervenience basis irreducible relations between particles occupying the singlet or triplet state, thus expanding the Humean mosaic. Darby claims that this move saves the reductive spirit of Humeanism, but he himself admits that some form of holism has to be accepted, which may be seen as departing from the Humean idea that everything that is is the local matters of particular fact. To that we would like to add that admitting irreducible, external relations between distinct individuals violates the intuition of no necessary connections. Clearly, the holding of the relation of anticorrelation between electrons in the singlet spin state implies that there is a necessary connection between the values of the spin of both electrons.}

In order to provide a philosophically clear discussion of this strategy, we will focus on the so-called ``primitive ontology'' approach to quantum physics \citep[see][for a technical discussion of the framework]{203}, In particular, we will follow Miller in framing the discussion in the context of non-relativistic Bohmian mechanics.\footnote{We stress the fact that here we are adopting a primitive ontology approach just as a working hypothesis that lets us exemplify the issues at stake, being fully aware that not everybody would agree with this choice (see \citealp{515}, for a critical discussion of the primitive ontology approach).}

\section{The Best System Approach to Bohmian Mechanics}\label{bsabm}

Bohmian mechanics, in rough outline, is based on the assumption that the world consists of fundamental particles which have well-defined trajectories. The modern version of the theory,\footnote{The roots of the theory can be traced back to the work of Louis de Broglie (see \citealp{338}).} as set out in \citet{323}, is in fact the simplest non-local Galilean-invariant theory of $N$ moving particles. The dynamics of the theory is given by:
\begin{subequations}\label{bm}
\begin{equation}\label{sch}
i\hbar\frac{\partial}{\partial t}\Psi(\mathbf{Q},t)=\hat{H}\Psi(\mathbf{Q},t);
\end{equation}
\begin{equation}\label{gui}
\frac{d\mathbf{Q}}{dt}=\hbar\mathbf{m}^{-1}Im\frac{\left<\Psi,\boldsymbol{\nabla}\Psi\right>}{\left<\Psi,\Psi\right>}(\mathbf{Q},t).
\end{equation}
\end{subequations}
$\mathbf{Q}=\left ( \begin{array}{l}
\mathbf{q}_{1}\\
\cdots\\
\mathbf{q}_{N}
\end{array} \right )\in \mathbb{R}^{3N}$ represents an instantaneous configuration of $N$ particles, $\boldsymbol{\nabla}=\left ( \begin{array}{l}
\boldsymbol{\nabla}_{1}\\
\cdots\\
\boldsymbol{\nabla}_{N}
\end{array} \right )$ is the ``gradient vector'', $\mathbf{m}$ is the $N\times N$ diagonal ``mass matrix'' $\{\delta_{ij}m_{i}\}$, and $\left<\cdot,\cdot\right>$ is an appropriate inner product defined over the space of wave functions.

To sum up, (\ref{gui}) is a concise way to write $N$ coupled equations of the form:
\begin{equation}\label{gui2}
\frac{d\mathbf{q}_{k}}{dt}=\frac{\hbar}{m_{k}}Im\frac{\left<\Psi,\boldsymbol{\nabla}_{k}\Psi\right>}{\left<\Psi,\Psi\right>}(\mathbf{Q},t).
\end{equation}

Formally, (\ref{gui}) depicts a vector field on $\mathbb{R}^{3N}$ depending on $\Psi$, whose integral curves $\mathbf{Q}=\mathbf{Q}(t)$ can each be ``unpacked'' as collections of $N$ continuous trajectories $\{\mathbf{q}_{i}=\mathbf{q}_{i}(t)\}_{i=1,\dots,N}$.\footnote{People like David Albert \citep[see, e.g.,][]{376} would deny the need for this last step. For them, the real dynamics literally unfolds in a higher dimensional space, where a single ``world-particle'' is pushed around by a ``$\Psi$-field''. We will not consider this controversial reading of \eqref{bm} here.} The dynamics encoded in (\ref{bm}) is deterministic: once provided a set of initial conditions $(\Psi_{0},\mathbf{Q}_{0})$ at a fixed time $t_{0}$, the dynamics singles out a unique dynamical evolution at earlier and later times. Moreover, it can be shown that (\ref{bm}) recovers Born's rule of standard quantum mechanics, thus matching all the empirical predictions of this latter theory (\citealp{222}, provide a detailed justification of this claim).

The above formalism establishes that Bohmian mechanics is a theory of $N$ point-like particles with definite positions $\mathbf{q}_{k}=(x_{k},y_{k},z_{k})$ in Euclidean $3$-space at all times. The non-locality of the theory is evident in (\ref{gui2}), the velocity of the $k$-th particle particle being instantaneously dependent on the positions of \emph{all} the other $N-1$ particles.  Furthermore, Bohmian mechanics is by construction a universal theory since (\ref{bm}) describes the dynamics of all there is in the universe --i.e. particles. The role of the wave function in this picture is to generate the vector field on the right-hand side of (\ref{gui}), and it is thus said to ``guide'' the motion of the particles. A key point to be stressed is that the formalism works well with \emph{any} type of wave function, including spinor states (of which singlet/triplet states represent a particular case).

Put in these terms, a natural interpretation of the wave function as a law-like element of the formalism suggests itself. To see this, we can follow (as people like Miller and Esfeld do) the well-trodden path taken by many Humeans with respect to the laws of nature. According to the Best System Approach (known also as the Mill-Ramsey-Lewis theory) laws are axioms systematizing our knowledge of individual facts that achieve the best balance between strength and simplicity (see \citealp{510}, for an exhaustive introduction to the subject). In other words, laws ultimately supervene on the collection of individual facts. In the same vein, the universal wave function can be treated as a law that gives us the best description of the behavior of individual particles.\footnote{However, see \citet{581} for a recent critique of the Best System approach to Bohmian mechanics. We do not wish to take a stand on whether this critique is serious enough to undermine Bohumianism independently of our objections developed in section \ref{se3}.} In this sense, the Humean treatment of quantum physics is very similar to that of classical physics. The mosaic is given by the history of change of particle positions\footnote{Standard Humeans would add some natural intrinsic properties, such as mass and charge, on top of positions. However, it is now clear that nothing over and above particles' (relative) positions and change thereof is needed in order to make sense of the Humean approach to laws of physics. See \citet{485} for a presentation of this ``Super-Humean'' stance.} over time and their mutual spatiotemporal relations, of which the wave function (or, say, the Hamiltonian, in the classical case) represent just (a part of) the simplest and most informative description. In this context, \emph{all} the physical properties usually encoded in the wave function can be reduced to the mutual arrangement of particles throughout the entire history of the universe, spin being one of such properties.\footnote{Such a construction is not immune to criticism. For example, \citet{617}, argues that purging the mosaic of all natural properties makes it impossible to establish which description among the many possible is in fact the simplest and most informative. \citet{618}, goes as far as arguing that a mosaic consisting only of material particles' trajectories is untenable.} It is important to stress the fact that this story makes sense because Bohmian mechanics by construction accords a metaphysically privileged status to \emph{positions}, otherwise one could wonder why, among the many physically possible bases onto which the quantum state could be projected, we just focus on the position basis. Hence, if we adopt this perspective, then the entanglement encoded in the wave function does not represent anymore a threat to a Humean reading of Bohmian mechanics. In fact, it seems that Bohmianism may be argued to satisfy Separability. To emphasize the Humean character of Bohmian mechanics, Miller even coins the term ``Bohumianism''.

It is enlightening to see how Bohumianism concretely defuses Maudlin's challenge. Let's start from the trivial case in which the two-particle system is all there is to the world. In this case, the mosaic would be too meager to be describable in terms of singlet/triplet states. Particles' relative motion might instead allow for, e.g., a description in terms of simple Coulomb interactions. In this case, Maudlin's challenge is dodged rather than defused. The most interesting case involves, of course, the actual world.

The first question to be answered, then, is how Bohmian mechanics is able to describe the behavior of our pair of entangled particles, given that they represent just a small portion of the universal $N$-particle configuration. Let's call $\mathbf{q}$ the two-particle subsystem and $\boldsymbol{\mathfrak{Q}}$ the rest of the configuration. Clearly, $\left(\mathbf{q},\boldsymbol{\mathfrak{Q}}\right)=\mathbf{Q}$. Now, in general there is no physically interesting case in which the universal wave function $\Psi$ can be written as a simple product state of the form $\Psi(\mathbf{Q})=\psi(\mathbf{q})\phi(\boldsymbol{\mathfrak{Q}})$. However, there are more physically interesting cases in which it can be written as $\Psi(\mathbf{Q})=\psi(\mathbf{q})\phi(\boldsymbol{\mathfrak{Q}})+\Psi^\perp(\mathbf{Q})$ \emph{and} the wave packet $\Psi^\perp(\mathbf{Q})$ remains empty throughout the dynamical evolution (i.e. it never gets to ``guide'' any particle).  This happens when the following conditions are met:

\begin{enumerate}
\item $\psi\phi$ and $\Psi^\perp$ do not substantially overlap in configuration space.
\item $\boldsymbol{\mathfrak{Q}}$ always lies in a region of configuration space where $\Psi^\perp$ is close to zero but $\phi$ is not.
\end{enumerate}

In this case, we recover the orthodox picture\footnote{The reader interested in nonstandard approaches to quantum measurements (especially those dispensing with decoherence), can take a look at, e.g., \citet{619}.} in which system and environment are assigned an \emph{effective} wave function, $\psi$ and $\phi$ respectively, each of which is subjected to a Schr\"odinger-like dynamics. From this point on, we can give the standard quantum description of $\psi(\mathbf{q})$ and $\phi(\boldsymbol{\mathfrak{Q}})$ in terms of a ``subsystem'' and its ``environment'' (see, \citet{222}, section 5, for a far more rigorous presentation of this topic).

The above story makes it manifest why there is strictly speaking nothing in the effective wave function $\psi$ that is intrinsic to the two-particle system. Indeed, $\psi$ is just the tip of the iceberg of a more complex description involving also the particles in the environment. In other words, we say that the two-particle subsystem is in a singlet or triplet state just in virtue of how the trajectories of the two particles are related to those of the particles making up, say, a Stern-Gerlach apparatus. In this sense, the appropriate ontological picture is not that of an initially isolated subsystem that gets mixed with the environment as a result of the dynamical evolution but, rather, that of a global mosaic of trajectories to which \emph{we} attach a simple and informative description in terms of subsystem/environment. The important point is that all the relevant information to account for, say, the detection of a two-particle system in a singlet or triplet state is crafted, so to speak, in the mosaic: nothing more is needed. In this sense, there is no question whether quantum physics is compatible with Humeanism.

Does all of this mean that Humeanism is fully vindicated in Bohmian mechanics? Well, not so fast!

\section{A Challenge to Bohumianism}\label{se3}
The key point of our critique of the above-described strategy is the observation that while Separability reflects strong Humean sentiments, it by no means exhausts the entire Humean doctrine, as encompassed in the two basic intuitions mentioned in section \ref{intr} (no irreducible modal facts and no necessary connections). In other words, we argue that it is possible to have a non-strictly Humean theory that still preserves Separability, and Bohmian mechanics may fall in this category. To see that, let us start by noting that there are two distinct ways of satisfying the requirement of Supervenience for a given physical system. One way is direct and straightforward: the total physical state of any system is intrinsic to it, and moreover supervenes on the Humean mosaic consisting entirely of the system's parts. However, Supervenience can also be satisfied vacuously, as it is essentially a conditional statement: if $P$ is an intrinsic property of $S$, then $P$ supervenes on the properties and spatiotemporal relations of $S$'s parts. But what if $S$ has no interesting intrinsic properties (i.e. its physical state turns out to be an extrinsic property)? Then the conditional is true, but its truth follows simply from the falsity of the antecedent. There is no genuine supervenience here, because there is no property to which the supervenience could be attributed in the first place.

The distinction we've made immediately suggests a natural strengthening of the Supervenience principle: rather than limiting the supervenience property to the intrinsic features of physical systems, we may extend it to all their physical states. Thus a stronger Supervenience presents itself:

\begin{quote}
(Strong Supervenience) For any physical system $S$, the complete physical state of $S$ supervenes on the intrinsic physical states and spatiotemporal relations between points and pointlike objects that are parts of $S$.\footnote{\citet{482} call this principle \emph{Strong Separability}.}
\end{quote}

It should be noted that the only difference between Strong Supervenience and Supervenience is that the former drops the condition of intrinsicality with respect to the complete physical state of $S$. However, this difference has some dramatic consequences. In particular, Bohmian mechanics violates Strong Supervenience, since the singlet/triplet states of two electrons do not supervene on the properties of the individual electrons and their spatiotemporal arrangements, as we have explained in the previous section. But is Strong Supervenience a necessary part of any Humean doctrine? Can a Humean accept a situation in which the total physical state of a system contains an irreducible reference to objects spatiotemporally separated from this system? We believe not, since such a scenario clearly flies in the face of the second Humean intuition regarding the lack of necessary connections between separate entities. It seems that a committed Humean should subscribe to the following variant of Separability that we call \emph{Binary Separability}, which entails Strong Supervenience:

\begin{quote}
(Binary Separability) For any physical system $S$ equipped with its own physical state, the complete state of the world supervenes on the intrinsic properties of $S$, intrinsic properties of its environment $E$, and the spatiotemporal relations between $S$ and $E$.
\end{quote}

According to Separability, the supervenience basis for the total state of the world consists of the \emph{most fundamental} elements of reality (points or point-like objects). Binary Separability, on the other hand, insists that a supervenience basis can be also found on non-fundamental levels. This stronger principle ensures that each time when we identify a physical system that is a proper part of the world (but does not have to be a simple element with no further proper parts), we can split the state of the world into two components: one that is associated with the selected system, and the other characterizing its environment (plus the requisite spatiotemporal relations between the two). In other words, physical systems do not enter into any relations with their environments that would not be already included in their intrinsic properties plus spatiotemporal arrangements.

That Binary Separability implies Strong Supervenience can be proven as follows. First, we can observe that Binary Separability implies Separability, since Binary Separability can be applied to fundamental systems (points or point-like objects), which proves that these systems and their intrinsic states belong to the supervenience basis for the entire world. Next, assume that Binary Separability is true and take any physical system $S$. Since the physical state of $S$ is part of the total state of the universe, by Separability it must supervene on the Humean mosaic of fundamental objects and properties. But if the state of $S$ supervened on fundamental objects that are not parts of $S$, this would violate Binary Separability, since the state of $S$ would no longer be an intrinsic property of $S$. Thus Strong Supervenience has to be true as well. Hence, if we can argue that Humeans should accept Binary Separability, they should also commit themselves to Strong Supervenience.

In order to help the reader keep track of the logical relations among the array of metaphysical claims introduced and discussed in this paper, we present the following diagram of mutual logical dependencies among these claims, where arrows indicate the relation of logical entailment:
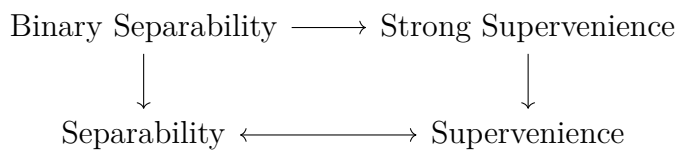
\begin{center}
\begin{tikzcd}[ampersand replacement=\&]
\text{Binary Separability} \arrow[r] \arrow[d]
\& \text{Strong Supervenience} \arrow[d] \\
\text{Separability} \arrow[r, leftrightarrow]
\& \text{Supervenience}
\end{tikzcd}
\end{center}
Our assertion is that the committed Humean should accept Binary Separability and thus all the remaining theses as well. However, even though Bohmian mechanics is compatible with Separability and Supervenience, it violates Binary Separability and Strong Supervenience, which poses a challenge to the Bohmians who want to subscribe to Humeanism.

Why isn't simple Separability enough for a Humean? Why do we need a stronger assumption regarding the supervenience of the whole on its parts? One reason may be the intuition that the relation of supervenience should ``mesh'' naturally with the intuitive mereological structure of the world.\footnote{Some might object that such a (standard) mereological structure is compatible with a \emph{classical} world. We are going to discuss this point in the next section.} Suppose that Binary Separability is false while Separability remains true. This means that the supervenience basis for the entire Humean mosaic for the world consisting of the fundamental objects and their properties cannot be divided up into the smaller supervenience bases for the objects composed of the corresponding elements of the fundamental mosaic. Let us choose a subset $\mathcal{P}$ of the set of all fundamental (point-like) objects that compose a particular physical system $S$. Even though the entire mosaic of point-like objects constitutes the supervenience basis for the world, the subset $\mathcal{P}$ does not analogously ground the state of system $S$. The state of system $S$ supervenes not only on $\mathcal{P}$ but on some elements outside of $\mathcal{P}$ as well. Thus the part-whole relation is not compatible with the supervenience relation.

Still, this argument may not convince all Humeans. For example, \citet{482} defend the view that Separability is all the Humean could ever want, and that introducing any stronger principle, such as Binary Separability or Strong Supervenience, does not add anything of value to the Humean. They believe that Separability already satisfies the requirements of no modal irreducible facts and of no necessary connections, so why bother? To that we can repeat what has already been said earlier that Separability assures only the non-existence of necessary connections between \emph{fundamental} entities, but does not guarantee that such connections will not appear at a higher level. Without Binary Separability a complex physical system may display irreducible connections with its environment due to the fact that the supervenience basis for its state extends beyond its spatiotemporal parts.\footnote{\citet{511} acknowledges that the acceptance of such connections amounts to the reinstatement of the anti-Humean doctrine of holism. She writes << [S]uppose it turns out that we can distinguish singlet and triplet pairs on the basis of differing relations their members bear to other elemental parts of the universe. We again might think this global interdependence itself indicates some kind of holism [...] regardless of whether entangled wholes bear any non-supervenient intrinsic properties>> (p. 512).}

In response to the above argument one may object that Humeans can and do happily accept some cases in which seemingly local states of affairs turn out to be dependent on the global distribution of properties.\footnote{We owe this objection to Michael Esfeld (private communication).} Take for instance the Humean regularity analysis of causation. Considering a single instance of causal interaction, for instance a stone smashing a window, the Humean will insist that this individual causal fact actually supervenes on the totality of similar cases in which a breakage of a fragile object follows a collision with a fast-moving projectile.\footnote{Another reductive analysis of that kind concerns individual chances of events which for Humeans are analyzable in terms of frequencies of occurrences of similar events throughout the history of the universe.} If the supervenience basis for a singular causal fact contains states of affairs external with respect to this fact, why can't the same apply to the case of the state of a physical system?

To that we reply that there is a fundamental ontological difference between purported causal facts and physical states of systems. Causality is an inherently modal notion, and as such is the subject of a reductionist analysis by Humean standards. Providing such a reductionist analysis in terms of global regularities does not show that a genuine local state of affairs supervenes on some extrinsic facts. Rather, Humeans would insist that singular causal facts are not local to begin with --they necessarily involve facts regarding external objects and systems. To put it differently, the Humean analysis of causation amounts to an elimination of causal connections between particular events, if we interpret such connections as modal facts intrinsic to appropriate pairs of events. But physical states of systems are supposed to be genuinely non-modal, so prima facie there is no reason why the Humean should reduce them to more fundamental properties and their distributions.

A similar response can be provided with respect to other purported counterexamples to the Humean prohibition of necessary connections between distinct entities and their properties. There is an undeniable necessary link between the value associated with a ten-dollar bill in my wallet and the decision of the Federal Reserve to increase the budget deficit by printing more money. Similarly, there a necessary connection linking the death of a husband to the change of the marital status of his wife from being married to being a widow. However, in both examples the properties involved (the purchasing power of a currency, the marital status of a person) are, to put it loosely, ``conventional" or ``human-dependent". And as such they are unlikely to be part of the most fundamental description of the world (they fail to ``carve nature at its joints'', using David Lewis' famous characterization of natural properties vis-\`a-vis non-natural ones; see e.g. \citealp{626}). In a sense, there is no objective, physical fact of the matter as to whether a person is married, divorced or widowed -- all there is here is a certain social norm that designates a person as such. And while Humeanism does not logically imply physicalism, it seems that in the context of the discussions on the metaphysics of fundamental scientific theories we should adopt physicalism as a working hypothesis. Thus we believe that whereas Humeans do not have to reject all necessary connections, they are certainly obliged to explain away such connections when they involve legitimate physical properties used in our best scientific theories, regardless of whether they are attributed to simple objects with no parts or complex entities. 

As we have shown, Bohumians are forced to accept the existence of necessary, nomological connections between the physical state of composite quantum systems and the configuration of the environment. We insist that this constitutes a violation of the general spirit of Humeanism, and we have tried to deal with some arguments purporting to show that Humeans can happily live with this form of necessary connections between distinct entities. At this point, however, we think that the onus is on the Bohumian to provide an explanation of why we are allowed to relax the strictures of traditional Humeanism as conceived by its founder, who definitely didn't have in mind the fundamental Bohmian particles when he scorned the existence of objective necessary links connecting distinct entities. In other words, why is the separability of the objects at the fundamental level enough for a modern Humean?

One legitimate way to respond to this challenge could be to insist that only fundamental entities exist in the literal sense of the word, while composite objects are mere constructs used for the purpose of economy and can be eliminated from the ontologically committing contexts.\footnote{We cannot offer any precise definition of the term ``literal existence'' (or ``fundamental existence'') other than that it is meant to refer to the irreducible ontological commitments of our best theories when expressed in a most parsimonious and simplest language possible, stripped of all metaphors and unnecessary vocabulary.} This position, akin to mereological nihilism, rejects the existence of necessary connections between complex entities simply by rejecting the existence of the entities in question. We may note that Bhogal and Perry do not subscribe to this view, since they explicitly include in the total state of the world what they call the L-state, i.e. the collection of physical states attributed to spatiotemporal regions that do not supervene on the states of their subregions \citep[][p. 77]{482}. Hence they have to assume the existence of objects (regions) that are neither point-like nor identical with the entire universe. This move can be seen as conflicting with the ontologically reductionist spirit of the Humean framework. However, at this point, it looks like the only way the committed Humean could abandon Strong Supervenience without reneging on the intuition of no necessary connections is by denying the literal existence of complex physical systems. This strategy will be the subject of the next section.


\section{A Possible Way Out}\label{se4}
So it seems that now the Bohumian is cornered. Either she accepts that Strong Supervenience is vacuously equivalent to Supervenience because there are no physical systems other than individual points (or point-like objects) and the universe, or she altogether abandons Strong Supervenience understood as a principle that is genuinely stronger than Supervenience, thus letting her framework be haunted by a ghost of necessity.

At this point, the Bohumian might be tempted to buy into the second horn of the dilemma, and just accept as a bare fact of the matter that some physical systems may display mutual irreducible connections that do not depend on their spatial separations. After all, Hume's tenets are a heritage from a pre-quantum era, so there would be nothing unreasonable in seeking to radically revise the doctrine under the light of modern physics. The Bohumian might further point out that the problem of necessary connections between distant systems plagues also Best System Approaches to classical mechanics \citep[see in particular][section 5]{422}. In that case, the problem is roughly that the way regularities in a small region $A$ of the mosaic determine inertial frames is so strong that it automatically fixes all the other inertial frames throughout the universe, including a far distant region $B$. Indeed, the construction makes automatically true the counterfactual ``had the regularities been different in $A$, the inertial frames in $B$ would have been different''.\footnote{Huggett is very careful to attach the notion of frame to that of a reference body (see \citealp{422}, page 46, second paragraph), so the talk of frames here is just a shorthand for referring to elements of the mosaic.} Going back to the quantum case, the Bohumian can furthermore point out that the existence of such irreducible necessary connections do not affect the way physics is done in the lab. It is in fact clear that for all practical purposes the effective wave function is insensitive to the exact configuration of the environment as long as no measurement-like interaction happens. So Bohumians may be tempted to downplay the ontological significance of the necessary connection between a given system and its environment. However, it is quite obvious why this FAPP approach fails. If we can convince ourselves that the quantum state of a system is practically intrinsic to this system, then Maudlin's original argument against Separability returns in full force!

In our opinion, the most promising way out of the impasse is to go for the first horn of the dilemma and show that this choice does not necessarily entail a commitment to any kind of anti-realism with respect to macroscopic objects radical enough to fly in the face of scientific practice (or even common sense, for that matter). Needless to say, this option requires a careful examination and discussion of its weak points, which we cannot afford in the remaining parts of this paper. We hope that our subsequent remarks, preliminary as they may be, will prompt further investigations by philosophers interested in these matters.

We first of all notice that quantum physicists themselves acknowledge the fact that entanglement is an ``infection'' that spreads via decoherence throughout the entire universe. This in particular means that even in standard physical practice the distinction between a quantum system and its environment is just an approximation made for practical purposes and valid on very small spacetime-scales. This is exactly what is conveyed in Bohmian mechanics through the treatment of effective wave functions presented in section \ref{bsabm}. Furthermore, the undivided nature of the universe is strongly implied by the universal non-local dynamics of the theory. These are the reasons why Bohmian mechanics does not support any part/whole relation that goes beyond the particle/configuration distinction.

So what is the place for electron pairs, or even tables, chairs, and the like in all of this? The key consideration here is that such a categorization of reality is made \emph{by us}, but may be argued to be unnecessary for Bohmian mechanics in order to be empirically adequate.\footnote{We do not wish to enter the debate on the exact meaning of empirical adequacy and its relation to perception and scientific practice. A proper analysis of these topics would require at least a separate article.}  All Bohmian mechanics needs in order to fit into the physical practice is presented in section \ref{bsabm}. But, if that is the case, why do we in fact make such a categorization? Because of the \emph{role} that these objects play in the lab\footnote{Here the word ``lab'' is intended in the broadest sense possible, including, e.g., astrophysical systems.} (or in our lives). Thus we call a thing ``chair'' because it has such and such shape \emph{and} it sustains our weight when we sit on it. Note that it is extremely simple to argue that the previous sentence is nothing but a shorthand description of how a bunch of particles trajectories relate to another bunch of particles trajectories throughout spacetime. Of course, it has to be stressed that what individuates a chair is its functional role, not just its shape, otherwise we could have cases in which a bunch of particles coalesce into ethereal chairs floating around. This is not to say that complex subsystems are mere arbitrary constructs. It is still the particle trajectories that determine which sub-configurations are salient and which are not. However, as already said, this salience boils down to the way particle trajectories are related. In this sense, there is no human arbitrariness involved in the definition of a chair, but still a chair is defined by the functional role it plays with respect to humans.\footnote{But not only humans, of course: a chair can sustain, say, a pile of books, not just a person!} That being said, it is also possible to go further and argue that our perception of a chair sustaining our weight can be given in terms of correlations between trajectories of ``chair'' particles and those of our ``brain'' particles. For those still skeptical that Bohmian mechanics supports such a functionalist-like reduction of subsystems we can point out that the procedure to define an effective wave function \emph{literally is} a functional definition of a subsystem.  Hence we submit that, by going for a functionalist-like account of subsystems, we can defend a minimalist Humean primitive ontology without claiming that such subsystems, including macroscopic objects, are mere illusions or arbitrary constructs.\footnote{This line of reasoning is adopted by \citet{508} as a reply to the challenge in \citet{507}.}

From what we have said so far, still it is not clear why Strong Supervenience and Binary Separability seem so compelling principles to our intuitions when, in fact, they cut no metaphysical ice. In other words, even granted that everything can be ontologically reduced in a strong sense to an underlying universal dance of particles, still there is no convincing explanation for the fact that it is totally natural to think (i) that all that can be predicated of a chair is reducible to facts happening in the spacetime region where the ``chair'' particles' trajectories are located, and (ii) that a coarse-grained version of the mosaic still represents a legitimate Humean supervenience basis. The answer to these doubts lies in the quantum-to-classical transition. Actually, the classical limit of Bohmian mechanics is still largely work-in-progress \citep[see, e.g.,][]{509}, so a full answer is still to come. Very roughly speaking, because of the decoherence mechanism, effective Bohmian states $(\psi,\mathbf{q})$ evolve at large spatiotemporal scales as classical states $(\mathbf{p},\mathbf{q)}$, thus washing away the non-local dependencies at a macroscopic level. This gives the appearance of having a situation in which Strong Supervenience holds, that is, where non-fundamental levels have the same ontological ``dignity'' as the fundamental one. Of course, overlooking this fact, and thus considering coarse-grained classical states in parallel with the fine-grained Bohmian dynamics would muddle the ontological waters enough to create the problems introduced in the previous section.


To sum up, the suggested strategy for Bohumians is to question the fundamental ontological reality of complex physical systems other than the entire universe. This move solves all the problems described earlier at one bold stroke. The theses of Separability and Supervenience become trivially equivalent, as now the universe is considered the only complex physical system in existence. Even more importantly, Supervenience and Strong Supervenience collapse into one, since the physical state of the universe is by definition intrinsic to its bearer. The success of this strategy, which may be called \emph{Austere} Bohumianism, depends primarily on the plausibility of the aforementioned story that purports to explain the appearances of complex structures, including but not limited to the macroscopic objects of our experience. Even if we accept that austere Bohumians can offer a convincing explanation of how such objects emerge from the underlying reality of swarming elemental particles, still one ontological price to be paid is the rejection of the principles of mereology. Humeanism is definitely not a free lunch.


\section{Conclusion}\label{seco}
In this paper we have tried to assess the compatibility of the Humean tenets with Bohmian mechanics. While we side with Miller, Esfeld, Bhogal, and Perry in claiming, pace Maudlin, that quantum mechanics does not put the \emph{reductionist core} of the Humean doctrine in serious jeopardy, we nonetheless submit that the literature on the subject has so far glossed over a serious consequence that salvaging Humeanism in a Bohmian framework implies. In fact, if we want to repel Maudlin's challenge, we basically need to abandon the intuitive --and scientifically successful-- picture of the world as having a standard, well behaved mereological structure ((in the sense of displaying a part-whole structure that meshes with the relation of supervenience, as explained in section \ref{se3}). The point, then, is to decide how much of the Humean doctrine actually rests upon the assumption of such a structure, that is, how much of the strong supervenience principle is dispensable without perverting the nature of Humean supervenience thesis. And even if Strong Supervenience can be salvaged in its entirety, the question remains what other concessions have to be made in order to preserve the main tenets of Humeanism.

In order to forestall possible objections, we would like to stress that we do not insist that there is one, well-defined view that can be properly called ``Humeanism''. On that issue we side with \citet{616} who wrote: ``[``Humean Supervenience''] is a name shared by many different theses, differing from one another in subtle ways, though they are all intended to capture the same general view of the world'' (p. 2), and then added even more forcefully ``Thus, HS [i.e. Humean Supervenience] has acquired a status like the one that doctrines like materialism, dualism, and empiricism often appear to have: there seems to be an idea there, that one can be determinately for or against, even while it remains an open question exactly how the idea should be formulated'' (p. 3). As we have explained above, we believe that one legitimate way to formulate the idea of Humean Supervenience is in terms of the principles we call Strong Supervenience or Binary Separability, even though we acknowledge that some philosophers who subscribe to the broad doctrine of Humeanism may be unwilling to accept these principles. By showing that Bohumianism does not respect Strong Supervenience or Binary Separability we do not commit a straw man fallacy against those philosophers, but rather bring to the surface a potentially troubling consequence of their position (the existence of necessary connections between complex physical systems), which in our opinion has been ignored in recent debates. If they are happy with a version of Humeanism that accepts this consequence, we have nothing to say against that, other than that there are other variants of Humeanism on the market.\footnote{Incidentally, it is worth pointing out that Earman and Roberts themselves argue for a variant of Humeanism that seems to be very close to our interpretation based on Binary Separability. They namely interpret the Humean base to which all the facts should be reducible as consisting of non-nomic facts that can be outcomes of spatiotemporally finite observation or measurement procedures (p. 17), and they add that their formulation does not make reference to point-like entities (p. 20 ft. 29). So it seems that on their approach states of composite physical systems can be included in the Humean base.}

The doctrine we call Austere Bohumianism restores the Humean order at the fundamental level of reality. It dispenses altogether with glaringly anti-Humean phenomena such as holism and non-separability that, as we are being told, permeate the quantum world. Yet this victory has its price, as we have already conceded. The complex systems emerging from the underlying dance of particles must be interpreted as ontologically secondary (in comparison to the full-fledged reality of Bohmian particles), lest they become a threat to Humeanism either by exhibiting holistic, irreducible features, or by displaying suspicious necessary connections with their environments.\footnote{It is interesting to note that this provides the proponents of mereological nihilism with a new argument based on the foundations of physics, aside from the usual arguments from metaphysics.} It is rather interesting to observe that while the Bohumians gladly accept that for all practical purposes the world behaves as if consisting of isolated and well-individualized systems equipped with their own quantum states, they are not so keen to admit that for all practical purposes the world at the micro-level but not limited to fundamental particles looks decidedly anti-Humean, as evidenced by the cases of singlet- and triplet-spin states of composite systems of particles. This ``illusion of anti-Humeanism'' evaporates when we move to the fundamental level of reality, but so does the familiar picture of chairs, trees, and many-particle systems of textbook quantum mechanics. It seems that Humeans may be ultimately tempted to go all the way and insist, as Democritus would put it, that nothing exists except atoms and empty space; everything else is opinion.

\pdfbookmark[1]{Acknowledgements}{acknowledgements}
\begin{center}
\textbf{Acknowledgements}
\end{center}
Tomasz Bigaj acknowledges financial support of grant No. \linebreak 2017/25/B/HS1/00620 from the National Science Centre, Poland. 

Antonio Vassallo worked on the first draft of this paper while being at the University of Barcelona as a Juan de la Cierva Fellow. Hence, he gratefully acknowledges financial support from the Spanish Ministry of Science, Innovation and Universities, fellowship IJCI-2015-23321. The rest of his work on the paper has been carried out at the Warsaw University of Technology with financial support from the Polish National Science Centre, grant No. 2019/33/B/HS1/01772.

\pdfbookmark[1]{References}{references}
\bibliography{biblio}
\end{document}